# The Italian Summer Students Program at Fermi National Accelerator Laboratory and other US Laboratories.

**C. Luongo**[1]
*Istituto Nazionale di Fisica Nucleare – Sezione di Pisa*
*Largo B. Pontecorvo, 3, 56127 Pisa, Italy*
*E-mail:* `cluongo@pi.infn.it`

**E. Barzi**
*Fermi National Accelerator Laboratory*
*PO Box 500, Batavia IL, 60510-5011, USA*
*E-mail:* `barzi@fnal.gov`

**G. Bellettini**
*Istituto Nazionale di Fisica Nucleare – Sezione di Pisa,*
*Largo B. Pontecorvo, 3, 56127 Pisa, Italy*
***Fermi National Accelerator Laboratory*** *E-mail:* `giorgiob@fnal.gov`

**S. Donati**
*University of Pisa and Istituto Nazionale di Fisica Nucleare – Sezione di Pisa*
*Largo B. Pontecorvo, 3, 56127 Pisa, Italy*
*E-mail:* `simone.donati@pi.infn.it`

Since 1984 INFN scientists performing experiments at Fermilab have been running a two-month summer training program for Italian students at the lab [1]. In 1984 the program involved only a few physics students from the Pisa group, but it was later extended to other groups and to engineering students. Since 2004 the program has been supported in part by DOE in the frame of an exchange agreement with INFN and has been run by the Cultural Association of Italians at Fermilab (CAIF, [2])**.** In 2007 the Sant'Anna School of Advanced Studies (Pisa) established an agreement with Fermilab to share the cost of four engineering students each year. In the 34 years of its history, the program has hosted at Fermilab approximately 530 Italian students from more than 20 Italian universities and from some non-Italian universities[2]. In addition, in the years 2010-2018, with the support of INAF, ASI, and CAIF**,** 25 students were hosted in other US laboratories and universities. The Fermilab training programs spanned from data analysis to design and construction of particle detectors and accelerator components, R/D on superconductive elements, theory of accelerators, and analysis of astrophysical data. At the other US laboratories the offered training was on Space Science. In 2015 the University of Pisa endorsed the program as one of his own Summer Schools [3]. The interns are enrolled as Pisa students for the duration of the internship. They are required to write summary reports published in the Fermilab and University of Pisa web pages. Upon positive evaluation by a University board, students are acknowledged 6 ECTS credits. In 2017 a new initiative of the Oxford Summer School "MovingKnowledge17" [4] allowed three Fermilab summer students to be trained in neutrino physics to spend the month of July at Oxford to follow an advanced course on HEP.

---

[1]Speaker
[2] In 2015 the program was extended to accept non-Italian students.

The entire program is expected to expand further under CAIF management. An agreement has been signed between ASI and CAIF, for ASI to support yearly three two-months fellowships in US space science laboratories. The program is also part of the Outreach of the European Projects MUSE (H2020-MSCA-RISE-2015, GA 690835), NEWS (H2020-MSCA-RISE-2016, GA 734303) and INTENSE (H2020-MSCA-RISE-2018, GA 822185 in preparation).

In the following we inform on student recruiting, training programs, and final evaluation.

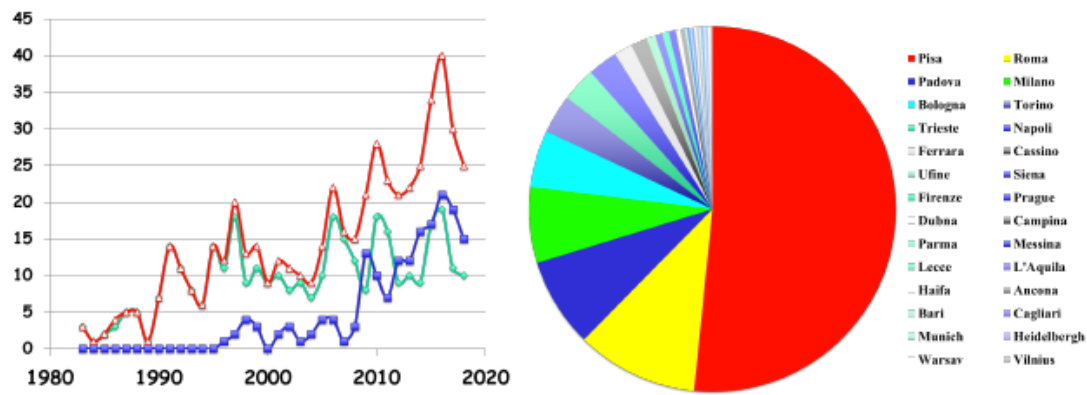

Figure 1 (left) Distribution of the number of students selected in the years 1984-2018: the blue curve represents the distribution of the engineering students, the green curve represents the distribution of the physics students and the red curve represents the total distribution. One notes an increased rate of engineering students in recent years, who are now as many as the physics students; (right) distribution of the number of students selected from each Italian and some European university.

## 1. Recruitment

In January we announce the available grants with posters and flyers, on the web and on most Italan Universities, whereby some basic information is given to the applicants[3]. The youngest undergraduates are excluded by requesting students of the "Laurea Magistrale" courses from Italian universities and of the Masters degree from European universities. Computer science skills and good knowledge of English are requested. In early spring CAIF members inquire about the available training programs at the lab, to be matched to the skills and interests of the trainees. One must find out which groups are interested in offering a training program to our students, and get a clear description of the offered program. It is essential to find a match between the profiles of the candidates and the interest of the groups. Only students matching an available training programs are ultimately selected. Pre-selected candidates are informed that they will have to discuss their case in person or by Skype with representatives of CAIF and of the sponsors. Thorough interviews are performed. The selected students are made known to supervisors as potential trainees in the months of April/May. The supervisors make their best choice and work offers are sent to the students. Finally, those students for whom a good fit and work program are found within a Fermilab team will enter the US with a J1 Visa for training in August and September. Free housing and shared rental cars are provided besides a weekly salary. Fermilab does not cover the round trip journey to the US and the health insurance. However, the salary is fully adequate to allow the students to cover the travel cost as well as all costs encountered in two months of stay. CAIF negotiates agreements with outside labs for training students sponsored by the Italian Space Science Agency (ASI, http://www.asi.it). As for the Fermilab students, early in the year the available ASI fellowships (currently 3) and the support offered by CAIF are announced, and in late spring the winners are selected. Their training period is from the end of July to the end of September. Figure 1 (left) shows the number of students selected each year from 1984 to 2018. Figure 1 (right) shows the students' sites of origin. About 75% of the students come from the universities of Pisa, Rome, Milan and Padova.

---

[3] As of now we have not been able to get this information distributed c/o the Italian Minister of University and Scientific Research. We shall continue shooting for this goal, which would allow for a much more efficient transfer of information to all Italian universities.

## 2. Logistics

The Fermilab Personnel Office receives the list of the selected students and e-mails them the job offers needed to get the J1 Visa required to be employed as Fermilab interns. Once the offer is accepted the students are responsible for getting their Visa. In addition, some paperwork is requested from the students by the Fermilab Visa Office. This includes accessing from Italy the "FermiWorks" website to complete an Onboarding process and filling a New Hire document. The rules for accessing the FERMI computing domain can also be followed from a distance, in order to gain access to the domain immediately upon arrival at the lab. At the student arrival CAIF members are around making sure that students housing proceeds smoothly in the lab dorm or in nearby hotels. On the first and the second day at the lab the students convene for an International Orientation session, where they are istructed on basic lab and US rules. A twin concluding session must be followed shortly before departure at the end of the training period. Although this bureaucracy may worry the students at the beginning, experience shows that they are able to learn quickly to fulfil all the requests, and that in a few days they are able to work efficiently in their groups, who are responsible for assigning to them adequate office space and computing power.

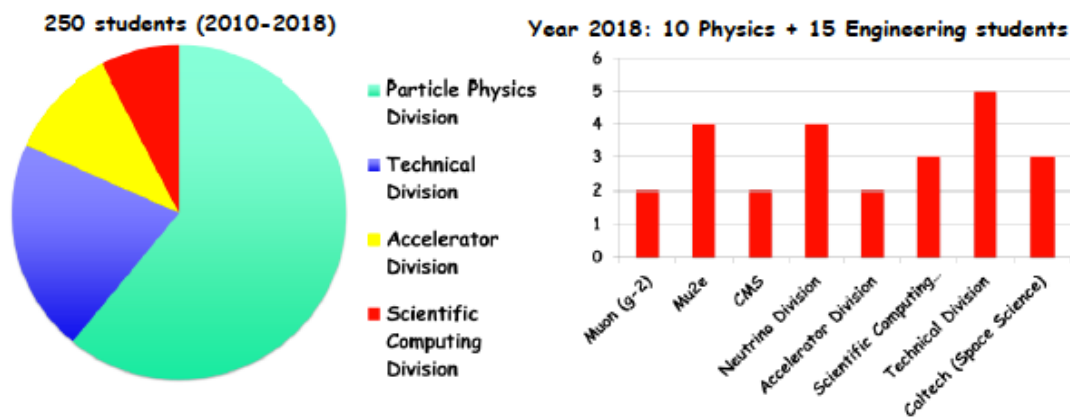

Figure 2: (left) Students' assignments among Fermilab Divisions in the years 2010-2018 and (right) among experimental groups in 2018.

## 3. Work Programs

The training programs span a very wide range of science and technology. The students are integrated in their research group and are encouraged to interact with as many colleagues as possible. The supervisors meet with their students on an individual basis, with meetings held at least once per week, to allow the group to profit of the productivity of the students. The students also participate in collaboration meetings, where they present their results in a wide professional environment. For physicists, students address analysis of experimental data in particle physics and astrophysics, simulation and setting-up of particle detectors, and particle accelerator theory. For engineers, they include fast digital electronics, design of detectors and accelerator components, superconducting materials and magnets, high precision mechanics, and advanced computing as well as civil engineering projects. Students make extensive use of advanced computation means sund programming languages, as C, C++, and Java, and apply advanced CAD and other technical tools for mechanics and electronics design (MatLab, OrCAD, Ansys, etc.). Physicists develop knowledge in statistical data analysis (Root). At Fermilab work is performed within projects, programs and experiments like Mu2e, Muon (g-2), NoVA, MicroBoone, LAriAT, Icarus, SBND, LBNF, DUNE, CDF, CMS, and General Accelerator R&D. Figure 2 (left) reports the distribution of the students' assignments among the Fermilab Divisions for the programs of the years 2010 to 2018. Figure 2 (right) shows the students' assignments to the Fermilab groups in 2018. Outside Fermilab, work is performed in astrophysics, space science and technology. Figure 3 shows the distribution of the 25 students among Nasa laboratories and US universities.

All students are requested to give midterm and final oral presentations and to write a technical report at the end of their stay. These documents are saved in the Fermilab Education Office web archive [5]. Upon successful completion of the final exam in front of a University of Pisa committee, students are acknowledged 6 ECTS credits. A significant fraction of former Summer Students have extended their collaboration with Fermilab performing Master and PhD Theses within the INFN groups or in other lab groups.

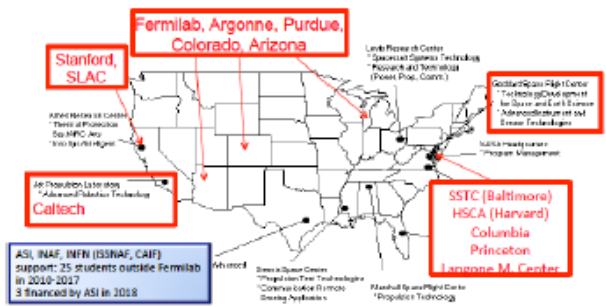

Figure 3: Space Science laboratories selected by the students for the INAF, ASI and INFN fellowships.

**Acknowledments**

This work was supported by the EU Horizon 2020 Research and Innovation Programme under the Marie Sklodowska-Curie Grant Agreement No. 734303.